\documentstyle[psfig,ltwol,psfig]{article}

\def\Journal#1#2#3#4{{#1} {\bf #2}, #3 (#4)}

\def\PLB{{\em Phys. Lett.}  B}
\def\PRL{\em Phys. Rev. Lett.}
\def\PRD{{\em Phys. Rev.} D}

\def\be{\begin{equation}}
\def\ee{\end{equation}}
\def\bea{\begin{eqnarray}}
\def\eea{\end{eqnarray}}

\def\nm{\mbox{$\nu_\mu$}}
\def\nmb{\mbox{$\bar{\nu}_\mu$}}

\begin{document}

\title{NEUTRINO TRIDENT PRODUCTION FROM NUTEV}

\author{
 T.~ADAMS$^{4}$$^{*}$, A.~ALTON$^{4}$, S.~AVVAKUMOV$^{7}$,
 L.~de~BARBARO$^{5}$, P.~de~BARBARO$^{7}$, R.~H.~BERNSTEIN$^{3}$,
 A.~BODEK$^{7}$, T.~BOLTON$^{4}$, J.~BRAU$^{6}$, D.~BUCHHOLZ$^{5}$,
 H.~BUDD$^{7}$, L.~BUGEL$^{3}$, J.~CONRAD$^{2}$, R.~B.~DRUCKER$^{6}$,
 R.~FREY$^{6}$, J.~FORMAGGIO$^{2}$, J.~GOLDMAN$^{4}$,
 M.~GONCHAROV$^{4}$,
 D.~A.~HARRIS$^{7}$, R.~A.~JOHNSON$^{1}$, S.~KOUTSOLIOTAS$^{2}$,
 J.~H.~KIM$^{2}$, M.~J.~LAMM$^{3}$, W.~MARSH$^{3}$, 
 D.~MASON$^{6}$, C.~McNULTY$^{2}$, K.~S.~McFARLAND$^{3,7}$, 
 D.~NAPLES$^{4}$, 
 P.~NIENABER$^{3}$, A.~ROMOSAN$^{2}$, 
 W.~K.~SAKUMOTO$^{7}$,
 H.~SCHELLMAN$^{5}$, M.~H.~SHAEVITZ$^{2}$, P.~SPENTZOURIS$^{2}$, 
 E.~G.~STERN$^{2}$, B.~TAMMINGA$^{2}$, M.~VAKILI$^{1}$, 
 A.~VAITAITIS$^{2}$, 
 V.~WU$^{1}$, U.~K.~YANG$^{7}$, J.~YU$^{3}$ and 
 G.~P.~ZELLER$^{5}$
}
\address{
 $^*$Presented by T. ADAMS \\ 
 $^1$University of Cincinnati, Cincinnati, OH 45221 \\            
 $^2$Columbia University, New York, NY 10027 \\                   
 $^3$Fermi National Accelerator Laboratory, Batavia, IL 60510 \\  
 $^4$Kansas State University, Manhattan, KS 66506 \\              
 $^5$Northwestern University, Evanston, IL 60208 \\               
 $^6$University of Oregon, Eugene, OR 97403 \\                    
 $^7$University of Rochester, Rochester, NY 14627 \\              
}


\twocolumn[\maketitle\abstracts{ 
A preliminary analysis of zero hadronic energy dimuon production in the 
NuTeV experiment
is presented ($\nm(\nmb) \ + \ N \ \rightarrow \ \nm(\nmb) \ + \ \mu^+ \ + \ 
\mu^- \ + \ N$).  The data show a signal in excess of background which is
attributed to neutrino trident production.  This excess is consistent
with the Standard Model prediction including electroweak interference.  
Other sources of low-$E_{HAD}$
dimuon events are also considered.
}]

\section{Introduction}

High energy physics has very few purely electroweak processes where
both $W^\pm$ and $Z^0$ contribute.  One such process is neutrino trident
production.

\begin{figure}
\center

 \begin{minipage}[tbp]{7.0cm}
  \psfig{figure=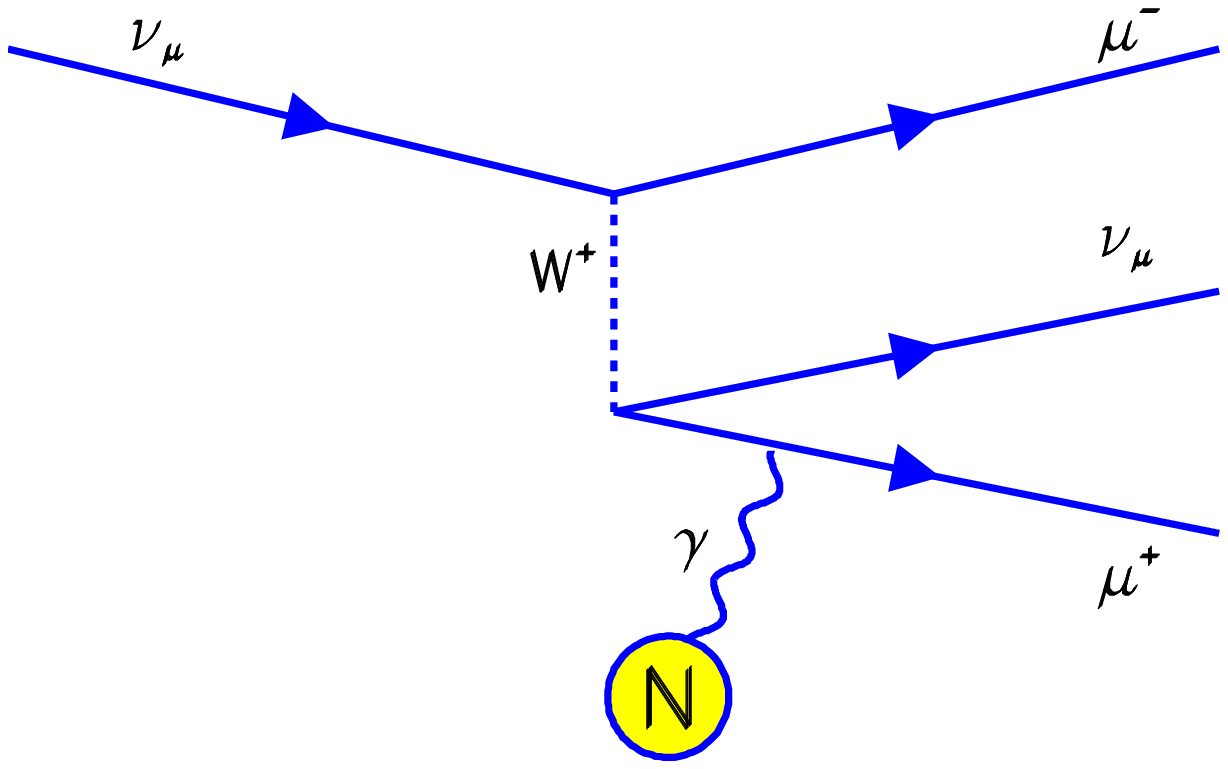,width=7.0cm}
 \end{minipage}
 \begin{minipage}[tbp]{7.0cm}
  \psfig{figure=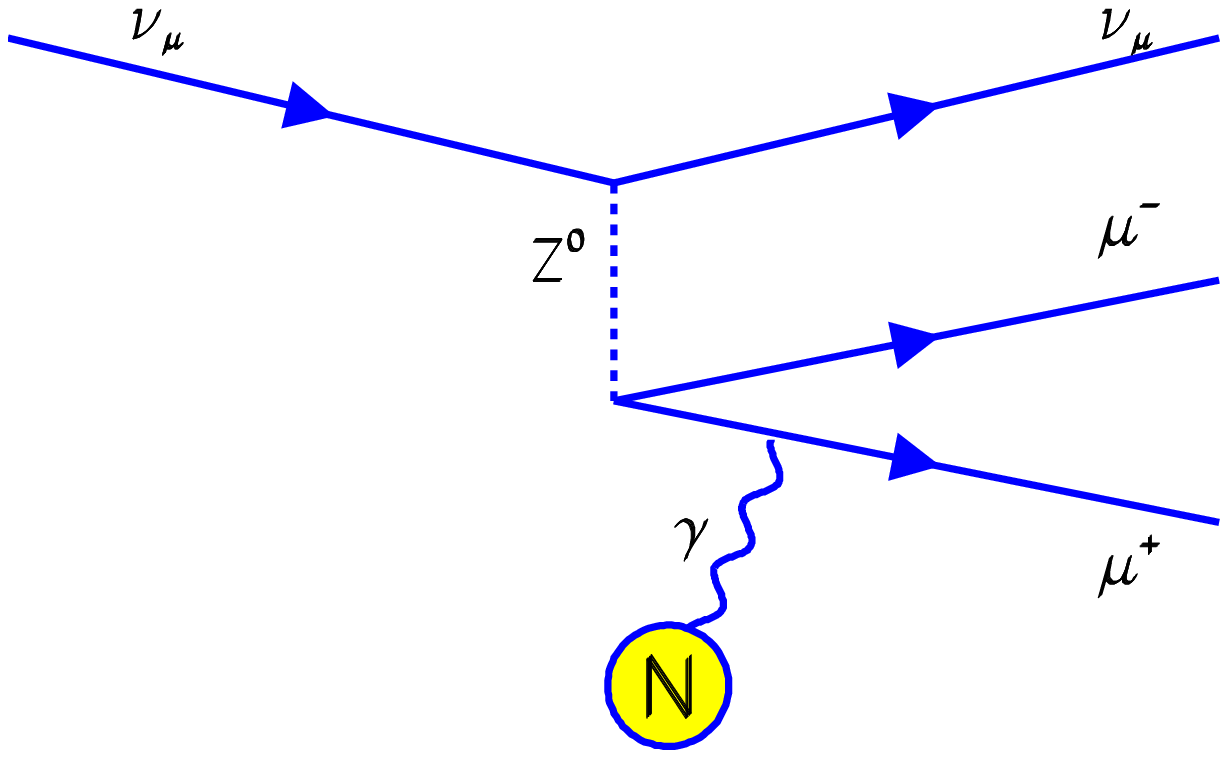,width=7.0cm}
 \end{minipage}
\caption{Feynman diagrams for neutrino trident production.}
\label{fig:feynman}
\end{figure}

Neutrino trident production is a purely electroweak process that
proceeds via exchange of a weak boson ($W^\pm$, $Z^0$) and 
a photon.  Feynman diagrams for two 
possible interactions are shown in Fig.~\ref{fig:feynman}. 

The reaction
\begin{equation}
\nm(\nmb) Z \ \rightarrow \ \nm(\nmb) \mu^+ \mu^- Z
\label{eq:trid}
\end{equation}
is of particular interest.  This is because it can proceed via 
either charged ($W^\pm$) or neutral ($Z^0$) currents 
(Fig.~\ref{fig:feynman}).
which results in interference.
Theoretical calculations~\cite{theory} predict this interference
to cause a 40$\%$ {\em decrease} in reaction~\ref{eq:trid}
from what would be expected for a purely charged-current 
interaction.  The measurement
of this interference is a test of the Standard
Model.

The final state in reaction~\ref{eq:trid} contains two muons
with no hadronic energy.  In a massive target experiment 
the muons make a clear signal by their penetration through the detector.
The muons should also have a very small opening angle and a small
invariant mass.
The dominant process producing two muons in the NuTeV
detector, neutrino charged-current DIS with a second
muon from decays of the final state hadrons, is
removed by requiring low energy deposition consistent
with two muons near the point of interaction.

Neutrino trident production has been investigated previously by several
experiments (CHARM, CHARM II and CCFR).  The extremely small cross-section 
($\sigma_{trident} = 10^{-5} \sigma_{CC}$)
makes it difficult to study.  The cross-section~\cite{theory} is proportional 
to

\begin{equation}
\sigma \propto \frac{2 Z^2 \alpha^2 G_F^2}{9 \pi^3} \frac{1}{R_{nucleus}}
E \log{E}.
\end{equation}

The CHARM experiment~\cite{charm}
failed to conclusively observe neutrino trident production. 
CHARM II~\cite{charm2} measured a
cross-section consistent with the Standard Model.  
Finally, CCFR~\cite{ccfr} ruled out a
pure $V - A$ description at the 99$\%$ confidence
level.  Each of these experiments used targets with different
Z's: CHARM (marble), CHARM II (glass) and CCFR (iron).

\section{Experimental Technique}

Reaction~\ref{eq:trid} can be studied by the NuTeV Experiment.
The following section describes the experiment, the technique
and issues related to the result.

The NuTeV Experiment (FNAL E815) uses the Fermilab Lab E detector and is 
very similar to that of its immediate predecessor, CCFR.~\cite{ccfr}
The primary components of the NuTeV detector are the 
target/calorimeter and the toroid spectrometer.

The target/calorimeter is a 690-ton steel sampling calorimeter.  It consists
of 42 segments which each contain 10-cm of iron, two
scintillation counters and one drift chamber.  The transverse dimensions
of each segment are 3 $\times$ 3 meters.  The energy resolution of
the calorimeter is $\sigma / E_{had} = 0.89 / \sqrt{E_{had}}$.

The toroid spectrometer is located immediately downstream of the
calorimeter.  It contains five sets of drift chambers in
an iron toroid magnet.  Momentum
measurement is limited by multiple scattering to $\sigma_p / p = 0.11$.
The magnetic field is set to focus the primary $\mu^-$($\mu^+$) from
\nm(\nmb) charged-current (CC) interactions.

The primary upgrade of the NuTeV experiment is the sign-selected
quadrapole train (SSQT) beamline.  This beamline provides sign-selection
of the secondary beam which produces a nearly pure \nm(\nmb)
beam.  The experiment ran in both \nm \ and \nmb \
modes.
This allows study
of \nm \ and \nmb \ trident production separately.

Selection criteria for this analysis are summarized in Table~\ref{tab:cuts}.
Events were selected with both muons momentum analyzed by
the toroid.  Muons were required to have a minimum energy 
($E_\mu$) and
have opposite charge.  The energy associated with a shower at the
vertex ($E_{HAD}$) must be very small.  The two muon
invariant mass ($M_{\mu\mu}$) is required to be small.
Finally, the fiducial volume is also limited to include regions
of the detector where the event could be well measured.

\begin{table}
 \begin{center}
 \caption{\label{tab:cuts} Selection criteria for neutrino tridents.}
 \vspace{0.2cm} 
 \begin{tabular}{|c|c|} \hline
  Variable & Criterion \\ \hline
  $E_\mu(1)$ (at event vertex) & $> 9$ GeV \\ \hline
  $E_\mu(1)$ (entering toroid) & $> 3$ GeV \\ \hline
  $E_\mu(2)$ (at event vertex) & $> 9$ GeV \\ \hline
  $E_\mu(2)$ (entering toroid) & $> 3$ GeV \\ \hline
  charge(1) + charge(2) & $= 0 $  \\ \hline
  \multicolumn{2}{|c|}{Both muon charges determined} \\ \hline
  $E_{HAD}$ & $< 3$ GeV \\ \hline
  $M_{\mu\mu}$ & $< 2.3$ GeV/c$^2$ \\ \hline
  Vertex Position & $> 12$ cm. from \\ 
  (Transverse) & calorimeter edge \\ \hline
  Vertex Position & $> 20$ cm of steel \\ 
  (Upstream) & from calorimeter edge \\ \hline
  Vertex Position ) &  $> 75$ cm of steel \\ 
  (Downstream) & from calorimeter edge \\ \hline

 \end{tabular}
 \end{center}
\end{table} 

A number of other physics processes can create a
signal similar to neutrino tridents.  A list of the most important
sources is given in Table~\ref{tab:back}.  The two sources with the
highest rate are charged-current charm
production and charged-current interactions with a $\pi / K$ decay
in the hadronic shower.  However, each of these processes generally
creates a significant amount of hadronic energy.  Monte Carlo studies
show that there is an expected background of $< 1$ event for both
of these combined.

Diffractive vector meson production can also produce a trident-like
signal.  Light vector mesons ($\rho^0$, $\omega$, $\phi$, and $J/\Psi$)
can be produced via neutral currents and decay into two muons.  
Charged-current production of $D_s^{+*}$ with a semi-leptonic decay ($D_s^+
\rightarrow \mu^+ + \nu_\mu + Y$) was presented at this 
conference.~\cite{chorus}  This may represent a significant background
to neutrino trident production.

Other small background sources include diffractive
$\pi^\pm$ production and  $\tau\mu$ trident production.  The
higher density target of CCFR/NuTeV limits the impact of diffractive
$\pi^\pm$ production compared to the lower target mass of CHARM II.
Monte Carlo studies will be performed to estimate 
the contributions from each of the background sources.

\begin{table*}
 \begin{center}
 \caption{\label{tab:back} Possible sources of background to neutrino
trident production. } 
 \vspace{0.2cm} 
 \begin{tabular}{|c|c|c|} \hline
  Source & Production & Decay \\ \hline 
  Charm Production & $\nu_{\mu} N \rightarrow \mu^-  c  X$ & $c 
\rightarrow \mu^+ \nu_{\mu}$ \\ \hline 
  $\pi / K$ decay & $\nu_{\mu} N \rightarrow \mu^- (\pi / K) Y$ & 
$\pi / K \rightarrow \mu^+ \nu_{\mu}$ \ or 
\\ 
  &  & $\pi / K \rightarrow \mu^+ \nu_{\mu} \pi^0$ \\ \hline 
  Vector Meson Production & $\nu_{\mu} N \rightarrow \nu_{\mu} V^0 \ X$   
   & $V \rightarrow \mu^+ \mu^-$ \\ 
  & ($V^0  = \{\rho^0, \omega, \phi, J/\psi\}$) & \\ \hline 
  $D_s^{+*}$ Production & $\nu_{\mu} N \rightarrow \mu^- D_s^{+*} X$ & $D_s^{+*} \rightarrow \gamma D_s^+ $ \\ 
  & & $D_s^+ \rightarrow \mu^+ \nu_{\mu}$ \\ \hline 
  $\pi^\pm$ Production & $\nu_{\mu} N \rightarrow \mu^- \pi^+ X$ & 
$\pi^+ \rightarrow \mu^+ \nu_{\mu}$ \\ \hline 
  $\tau\mu$ Trident Production & $\nu_{\mu} N \rightarrow \mu^- \tau^+  
\nu_{\tau} N$ & $\tau \rightarrow \mu^+ \nu_{\mu} \overline{\nu}_{\tau}$
\\ 
 \hline
 \end{tabular}
 \end{center}
\end{table*}

For the analysis presented here, the backgrounds are inferred from
the data rather than using Monte Carlo estimates.
The technique used is 
similar to that of the previous experiments.~\cite{charm,charm2,ccfr}
It is based on the following procedure.
The $E_{HAD}$ cut is removed and is
plotted for either side of one of the other criteria ($M_{\mu\mu} < 2.3$
GeV/c$^2$).
The distributions are normalized and the distribution for 
$M_{\mu\mu} > 2.3$ GeV/c$^2$
is used to estimate the background for 
$M_{\mu\mu} < 2.3$ GeV/c$^2$.

\begin{figure}
\center
\centerline{\psfig{figure=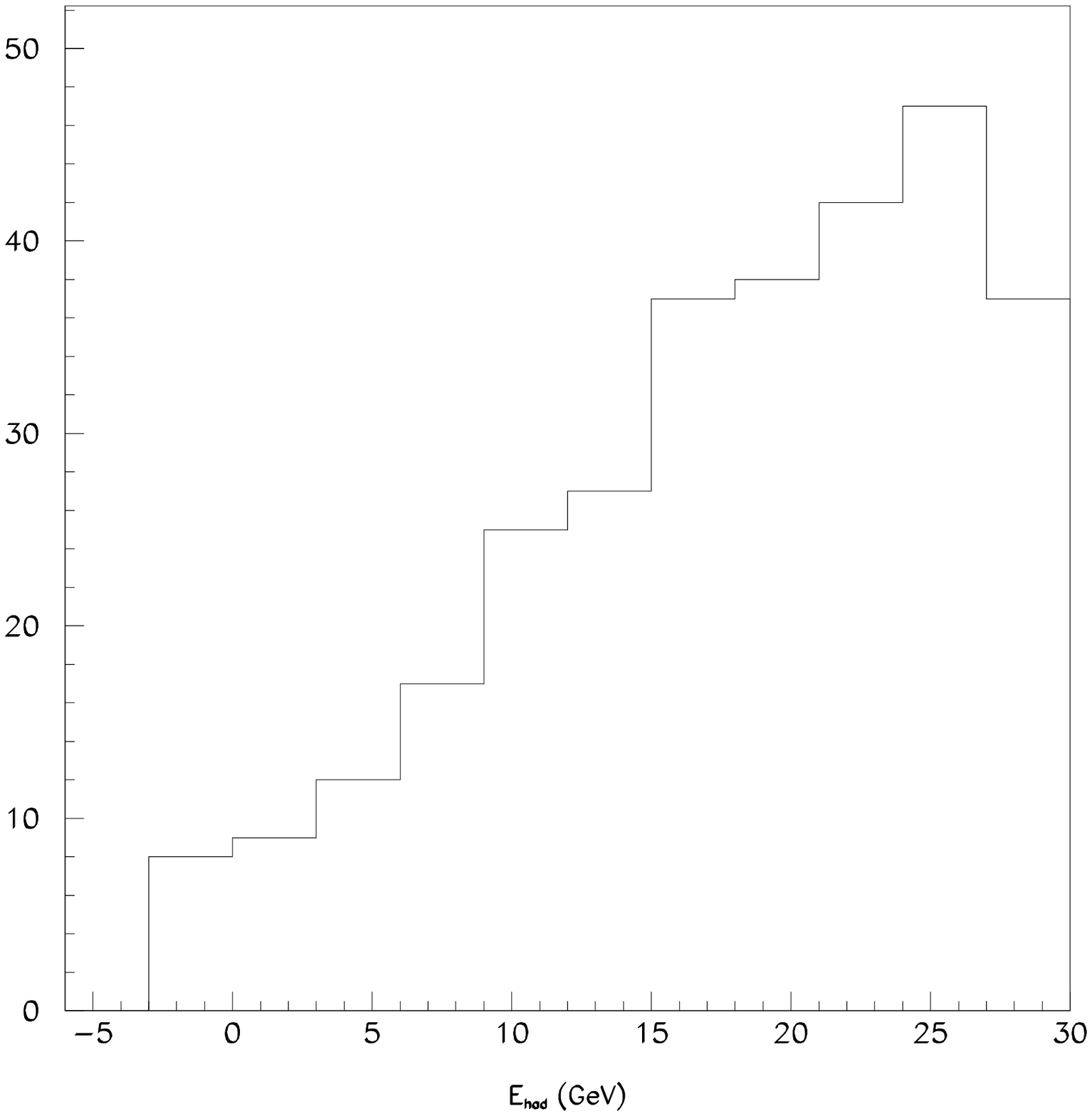,width=7.0cm}}
\caption{$E_{HAD}$ distribution for events with $M_{\mu\mu} < 2.3$ GeV/c$^2$.}
\label{fig:lowm}
\end{figure}

\begin{figure}
\center
\centerline{\psfig{figure=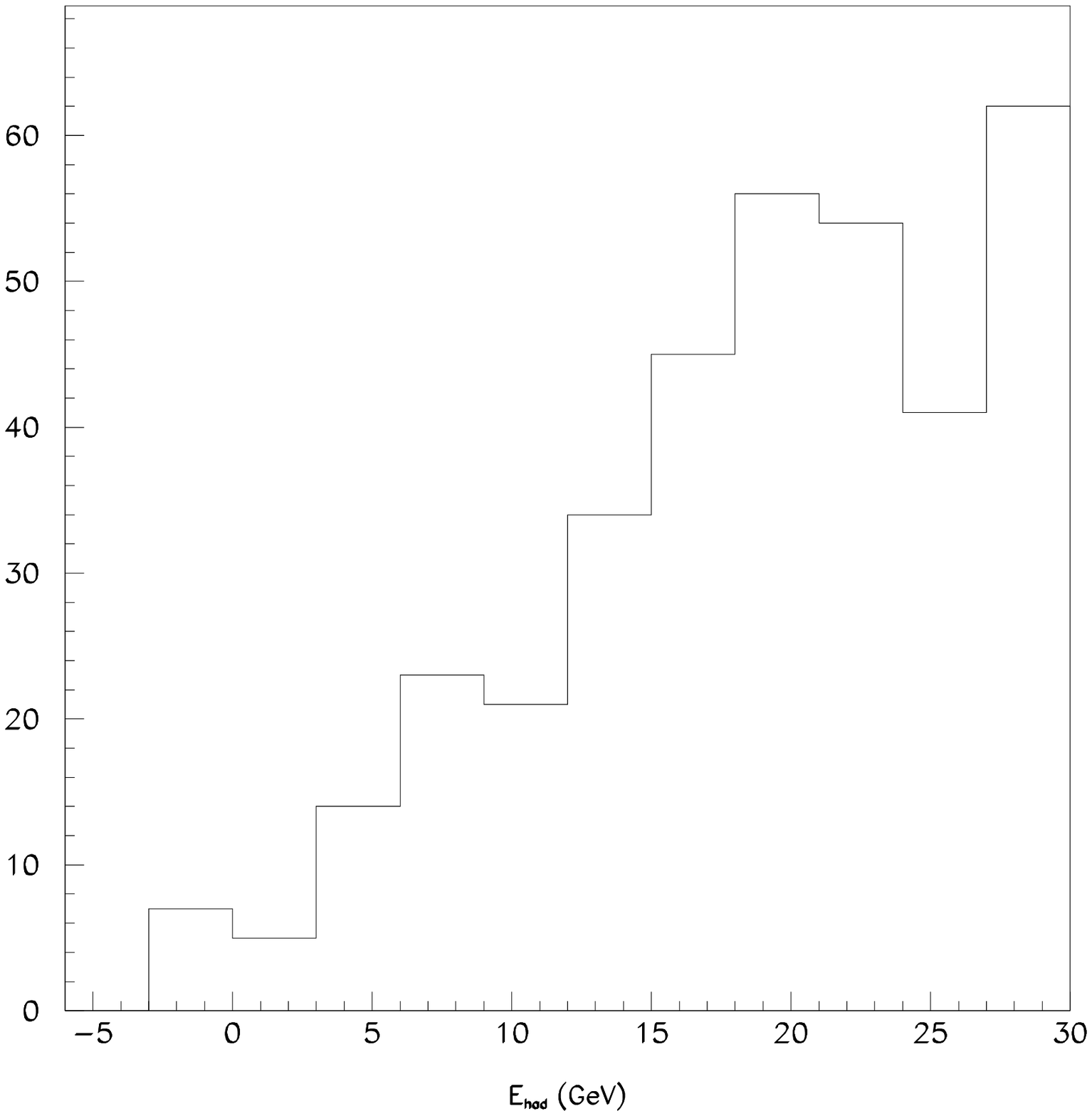,width=7.0cm}}
\caption{$E_{HAD}$ distribution for events with $M_{\mu\mu} > 2.3$ GeV/c$^2$.}
\label{fig:highm}
\end{figure}

\begin{figure}
\center
\centerline{\psfig{figure=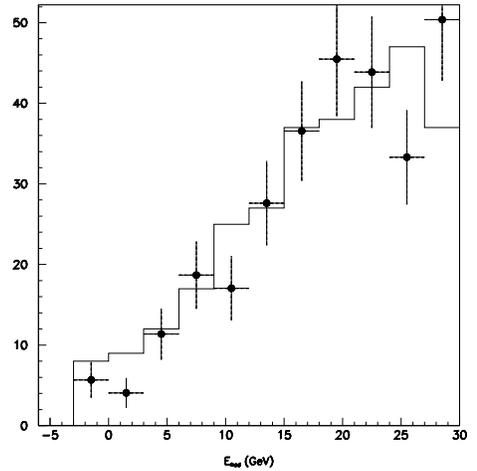,width=7.0cm}}
\caption{Normalized $E_{HAD}$ distribution for events with $M_{\mu\mu} < 2.3$ GeV/c$^2$ (histogram) and $M_{\mu\mu} > 2.3 GeV/c^2$ (points).}
\label{fig:bothm}
\end{figure}

Figures~\ref{fig:lowm}-\ref{fig:bothm} illustrate the background estimate.
All three plots show the $E_{HAD}$ distribution.  Figure~\ref{fig:lowm}
shows the distribution for events which pass the invariant mass cut
($M_{\mu\mu} < 2.3$ GeV/c$^2$).  Figure~\ref{fig:highm} shows the
distribution for events which fail the cut ($M_{\mu\mu} > 2.3$ GeV/c$^2$).
A normalization is performed to match the areas of the distributions
for $3.0 < E_{HAD} < 30.0$ GeV.  

The normalized distributions are plotted
together in Fig.~\ref{fig:bothm}.  The errors on the background estimate
the errors from the statistics of Fig.~\ref{fig:highm} and the error
on the normalization.  The lowest three bins ($E_{HAD} < 3.0$ GeV) are the
bins of interest for the neutrino trident signal.

\section{Results}

\begin{table*}[tb]
 \begin{center}
 \caption{\label{tab:results}Preliminary results for neutrino trident
analysis.}
 \vspace{0.2cm} 
 \begin{tabular}{|lcrrr|} \hline
  &  & Background & Standard & \\ 
  & Data & Estimate & Model~~ & V-A~~~~ \\ \hline
 $\nu$ mode & 12 & $7.6 \pm 2.5$ & $7.2 \pm 0.3$ & $12.26 \pm 0.5$ \\
 $\bar{\nu}$ mode & 5 & $2.0 \pm 1.4$ & $3.6 \pm 0.2$ & $6.1 \pm 0.3$ \\
 Combined & 17 & $9.8 \pm 2.9$ & $10.8 \pm 0.3$ & $18.3 \pm 0.6$ \\
 \hline
 \end{tabular}
 \end{center}
\end{table*}

A preliminary analysis of data from the 1997-98 run of NuTeV results 
in 17 events which pass
the trident selection criteria (12 in \nm, 5 in \nmb).  
Results are summarized
in Table~\ref{tab:results}.  The background estimate does not account for
all of the signal events and this excess is attributed to neutrino trident
production.

Table~\ref{tab:results} also shows the expected number of trident events
for both the Standard Model and $V - A$.  The neutrino trident excess is
consistent with the Standard Model and does not favor $V - A$.  This is
in agreement with CHARM II~\cite{charm2} and CCFR.~\cite{ccfr}

\section{Conclusions}

The NuTeV experiment has a sample of low-$E_{HAD}$ dimuon events.
A preliminary analysis has been performed which observes 
a signal consistent with neutrino trident production.  
Statistics will be improved by increasing acceptance 
and combining with data from CCFR.  This will yield the
highest statistics neutrino trident analysis to date.  Other 
sources of low-$E_{HAD}$ dimuons will be studied.

\end{document}